\documentclass[apj]{emulateapj}

\usepackage{color}
\usepackage{CJKutf8}
\usepackage[dvipsnames]{xcolor}
\usepackage[colorlinks=true,citecolor=blue,linkcolor=blue]{hyperref}

\usepackage{graphicx}
\usepackage{lineno}
\begin{document}       
      
\newcommand {\ha}{H$\alpha$}
\newcommand {\hb}{H$\beta$}

\newcommand {\kms}{km~s$^{-1}$}
\newcommand {\msun}{$M_\odot$}
\newcommand {\teff}{$T_{\rm eff}$}

%\begin{CJK*}{UTF8}{bsmi}

\title{New Insights on 30 Dor B Revealed by High-Quality Multi-wavelength Observations}

% The list of authors, and the short list which is used in the headers.
% If you need two or more lines of authors, add an extra line using \newauthor
\author{Wei-An Chen \begin{CJK}{UTF8}{bsmi}(陳韋安)\end{CJK}$^{1, 2}$,
Chuan-Jui Li \begin{CJK}{UTF8}{bsmi}(李傳睿)\end{CJK}$^2$, 
You-Hua Chu \begin{CJK}{UTF8}{bsmi}(朱有花)\end{CJK}$^2$,
Shutaro Ueda \begin{CJK}{UTF8}{bsmi}(上田周太朗)\end{CJK}$^2$, 
Kuo-Song Wang \begin{CJK}{UTF8}{bsmi}(王國松)\end{CJK}$^2$,
Sheng-Yuan Liu \begin{CJK}{UTF8}{bsmi}(呂聖元)\end{CJK}$^2$, 
Bo-An Chen \begin{CJK}{UTF8}{bsmi}(陳柏安)\end{CJK}$^3$
}

\affil{% List of institutions
$^{1}$ Graduate Institute of Astrophysics, National Taiwan University, No.\ 1, Sec.\ 4, Roosevelt Rd., Taipei 10617, Taiwan, R.O.C.\\
$^2$ Institute of Astronomy and Astrophysics, Academia Sinica, No.\ 1, Sec.\ 4, Roosevelt Road, Taipei 10617, Taiwan, R.O.C. \\
$^3$ Department of Physics, National Taiwan University, No.\ 1, Sec.\ 4, Roosevelt Rd., Taipei 10617, Taiwan, R.O.C.
}

%\author{Wei-An Chen \begin{CJK}{UTF8}{bsmi}(陳韋安)\end{CJK}}
%\affiliation{Graduate Institute of Astrophysics, National Taiwan University, No.1, Sec.4, Roosevelt Rd., Taipei 10617, Taiwan, R.O.C.}

%\author{Chuan-Jui Li \begin{CJK}{UTF8}{bsmi}(李傳睿)\end{CJK}, You-Hua Chu \begin{CJK}{UTF8}{bsmi}(朱有花)\end{CJK}, Shutaro Ueda \begin{CJK}{UTF8}{bsmi}(上田周太朗)\end{CJK}, Sheng-Yuan Liu \begin{CJK}{UTF8}{bsmi}(呂聖元)\end{CJK}}
%\affiliation{Institute of Astronomy and Astrophysics, Academia Sinica, No.\ 1, Sec.\ 4, Roosevelt Rd. Taipei 10617, Taiwan, R.O.C.}

%\author{Bo-An Chen \begin{CJK}{UTF8}{bsmi}(陳柏安)\end{CJK}}
%\affiliation{Graduate Institute of Astrophysics, National Taiwan University, No.1, Sec.4, Roosevelt Rd., Taipei 10617, Taiwan, R.O.C.}

%======================================================================
% abstract
%======================================================================
%

\begin{abstract}
The supernova remnant (SNR) 30 Dor B is associated with the \ion{H}{2} region ionized 
by the OB association LH99. The complex interstellar environment has made it difficult 
to study the physical structure of this SNR. We have used Hubble Space Telescope H$\alpha$ 
images to identify SNR shocks and deep Chandra X-ray observations to detect faint diffuse 
emission.  We find that 30 Dor B hosts three zones with very different X-ray surface 
brightnesses and nebular kinematics that are characteristic of SNRs in different interstellar 
environments and/or evolutionary stages.  The ASKAP 888 MHz map of 30 Dor B shows 
counterparts to all X-ray emission features except the faint halo. The ASKAP 888 MHz and 
1420 MHz observations are used to produce a spectral index map, but its interpretation is 
complicated by the background thermal emission and the pulsar PSR J0537$-$6910's flat spectral index. 
The stellar population in the vicinity of 30 Dor B indicates a continuous star formation 
in the past 8--10 Myr.  The observed very massive stars in LH99 cannot be coeval with the 
progenitor of 30 Dor B's pulsar.  Adopting the pulsar's spin-down timescale, 5000 yr, as 
the age of the SNR, the X-ray shell would be expanding at $\sim$4000 km\,s$^{-1}$ and the 
post-shock temperature would be 1--2 orders of magnitude higher than that indicated by the 
X-ray spectra.  Thus, the bright central region of 30 Dor B and the X-ray shell requires 
two separate SN events, and the faint diffuse X-ray halo perhaps other older SN events.
\end{abstract}

\subjectheadings{ISM: supernova remnants --- ISM: individual objects (30 Dor B) --- Magellanic Clouds}

%======================================================================
\section{Introduction}  \label{sec:Introduction}
%======================================================================

The name ``30 Dor B'' originates from the first 408 MHz map that resolved 
the 30 Doradus (30 Dor) region into A, B, and C components \citep{LeMarne1968},
corresponding to the giant \ion{H}{2} region centered on the OB association 
LH100 (aka the R136 cluster), the \ion{H}{2} region of the OB association LH99, 
and the superbubble around the OB association LH90, respectively \citep{Lucke1970}. 
These three components not only are spatially distinct, but also have different 
stellar population and nebular morphology.  
The A component has been simply called 30 Dor, while the other two components 
are commonly called 30 Dor B and 30 Dor C.  
The high radio-to-optical emission ratio of 30 Dor B led to the suggestion of 
a supernova remnant (SNR), and its flat radio spectrum prompted the speculation 
that the SNR might be Crab-like \citep{Mills1978}.
The detection of diffuse X-ray emission confirmed the existence of an SNR in 
30 Dor B \citep{Long1981}.
Strictly speaking, the name ``30 Dor B" refers to the region that contains both 
the \ion{H}{2} region of LH99 and the SNR, while the SNR per se should be called 
MCSNR J0537$-$6910; however, to keep the name short, we will continue calling the 
SNR ``30 Dor B" in this paper.  As 30 Dor has been cataloged as LHA 120-N157
\citep{Henize1956}, or N157 for short, 30 Dor B has also been called N157B.

The position-coincidence of 30 Dor B and the \ion{H}{2} region photoionized 
by the OB association LH99 \citep{Lucke1970} points to a probable physical association,
thus its SN progenitor is suggested to be a massive member of LH99 \citep{Chu1997}.
30 Dor B has been observed in essentially all wave bands.  
Its radio observations made with the Australia Telescope Compact Array
at wavelengths of 3.5 and 6 cm indicate a radio spectral index of $-$0.19
in the pulsar wind nebula (PWN), named PWN N157B \citep{Lazendic2000}. 
Spitzer IR observations of 30 Dor B reveal mostly thermal emission from 
dust heated by radiation of massive stars in LH99 \citep{Micelotta2009}.
High-dispersion optical spectroscopic observations of 30 Dor B have been 
used to kinematically identify SNR shocks and boundary \citep{Chu1992}.
Rossi X-Ray Timing Explorer observations detected the 16 ms pulsed X-ray
emission from the pulsar PSR J0537$-$6910 \citep{Marshall1998} and 
Chandra X-ray observations resolved its PWN \citep{Wang2001}.
Even high-energy gamma rays have been detected from this PWN 
in 30 Dor B \citep{HESS2012}.

While 30 Dor B is distinct from the main body of the 30 Dor giant \ion{H}{2} 
region, it has been included in the Hubble Tarantula Treasury Project 
(HTTP, PI: E.\ Sabbi) and high-resolution Hubble Space Telescope (HST) 
continuum and H$\alpha$ images of 30 Dor B become available.  
These images clearly reveal sharp filaments associated with the SNR shocks.  
30 Dor B is also included in the Chandra X-ray Observatory's very large program 
The Tarantula - Revealed by X-rays (T-Rex, PI: L.\ Townsley), which has mapped
30 Dor with ACIS-I in 2 Ms. 
This deep observation reveals the full extent of diffuse X-ray emission 
associated with 30 Dor B and clearly shows, in addition to the PWN, 
a shell structure and a faint halo.  
Now is thus an opportune time to re-examine 30 Dor B
and make a comprehensive multi-wavelength analysis of its structure and 
relationship with the ambient interstellar medium (ISM).  
We have retrieved high-resolution multi-wavelength images available in archives, 
evaluated the effectiveness of the SNR diagnostics, and assessed the physical
structure of the 30 Dor B SNR. 
This paper reports our analysis and results.

%======================================================================
\section{Observations} \label{sec:Observations}  
%======================================================================

\begin{figure*}
\plotone{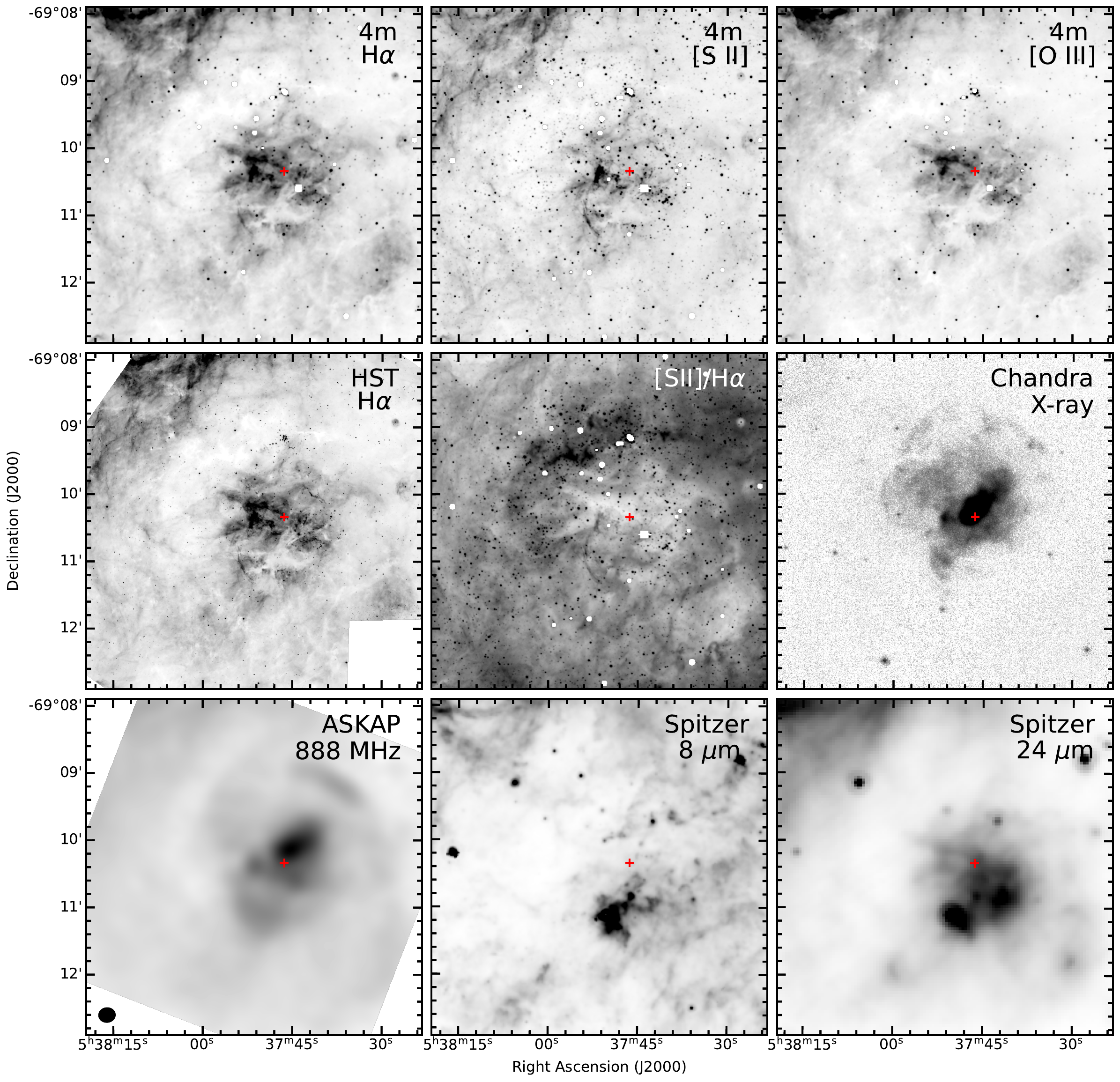}
\caption{Multi-band images of 30 Dor B. All images use "inverted colormap" with high values 
in black and low values in white.  The top row shows the H$\alpha$, [\ion{S}{2}] and [\ion{O}{3}] images
taken with the MOSAIC camera on the Blanco 4m telescope. The white spots are bright stars that are 
saturated in the MOSAIC images.  The middle row shows the H$\alpha$ image from HST, 
[\ion{S}{2}]/H$\alpha$ ratio from 4m MOSAIC images (see Fig.~\ref{fig:sii_ha_raio} for ratio maps with 
color bar), and X-ray 0.5--7.0 keV band image (log scale) from 2.18 Ms Chandra ACIS observations. 
The bottom row shows the radio image at 888 MHz from ASKAP (square root scale, with the beam plotted at the lower left corner), 
IRAC 8 $\mu$m and MIPS 24 $\mu$m images from Spitzer.  The red ”+” symbol in each panel marks
the position of the pulsar PSR J0537$-$6910.}
\label{fig:multi_band}
\end{figure*}

\begin{figure*}
\plottwo{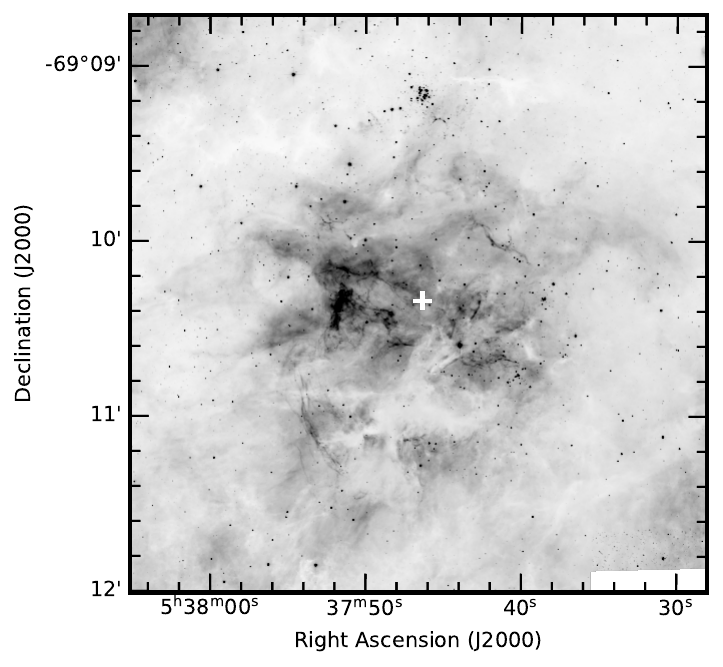}{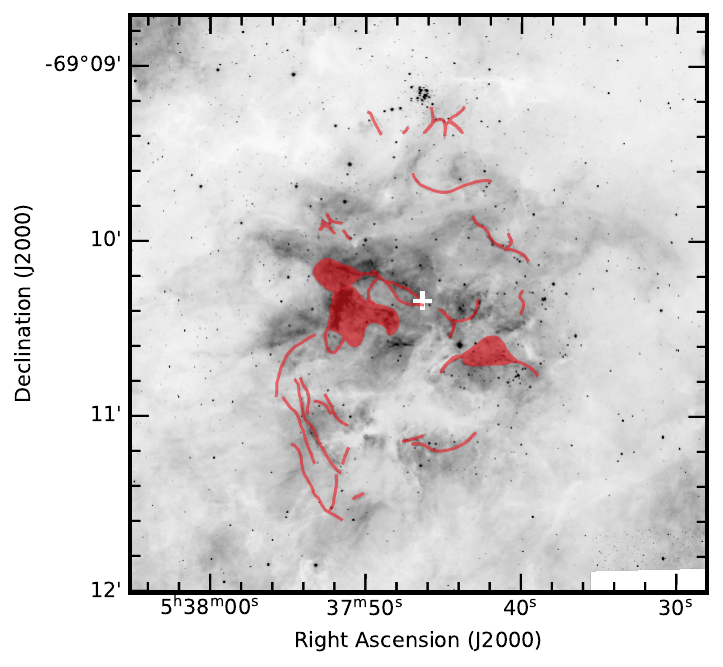}
\caption{Left: Closeup HST H$\alpha$ image of 30 Dor B. Right: HST H$\alpha$ image with
sharp filaments marked. The white "+" symbol marks the position of the pulsar PSR J0537$-$6910.}
\label{fig:30DorB_filaments}
\end{figure*}

%\subsection{HST Images}

We use archival HST mosaic images of 30 Dor from the 
HTTP \citep{Sabbi2013, Sabbi2016, SabbiElena2016}
to study the 30 Dor B SNR and its interactions with the ambient ISM. 
The HTTP is a panchromatic imaging survey of stellar populations in the 
Tarantula Nebula (aka 30 Dor). The HST mosaic images used observations
made with the Wide Field Channel of the Advanced Camera for Surveys 
(ACS/WFC), the UVIS channel of the Wide Field Camera 3 (WFC3/UVIS), and 
the IR channel of the Wide Field Camera 3 (WFC3/IR) from mainly 
Program 12939 (PI: E.\ Sabbi) and partly Program 12499 (PI: D.\ Lennon).
For these mosaic images, the pixel size is 0\farcs04, corresponding to 0.01 pc 
in the Large Magellanic Cloud (LMC, at a distance of 50 kpc), and 
the orientation (the up direction) is 35$\degr$ clockwise from 
the North. The available HST mosaic images from the HTTP are listed in 
Table \ref{table:hst_observations}. 
The HTTP data product described above may be retrieved from the MAST 
archive through
\dataset[doi:10.17909/T9RP4V]{https://dx.doi.org/10.17909/T9RP4V}.

% The exposures were astrometrically aligned to a master reference catalog derived 
%from the Gaia Data Release 1, selecting only sources with positional uncertainties 
%less than 5 mas.

%The science products (sci) are in units of electrons/s and are accompanied by 
%weight files (wht) which are based on the exposure time of all the input exposures.

%--------------------------------------------
%             Table 1  
  
\begin{deluxetable}{ccccc}
\tabletypesize{\scriptsize}
%\rotate
\tablewidth{0pc}
\tablecaption{HST Observations}
\tablehead{
Program ID & PI & Instrument  & Filter }
\startdata 
 GO-12499   & D. Lennon & ACS/WFC &  F775W \\
  . &  .  & WFC3/UVIS & F775W  \\
\hline
 GO-12939  &  E. Sabbi &  ACS/WFC &  F555W \\
 .  &  . & ACS/WFC &  F658N \\
 .  &  . & WFC3/IR &  F110W \\
 .  &  . & WFC3/IR &  F160W \\
 .  &  . & WFC3/UVIS  &  F275W \\
 .  &  . & WFC3/UVIS &  F336W 
\enddata
\tablecomments{See \cite{Sabbi2013, Sabbi2016} for further details about the observations.}
\label{table:hst_observations}
\end{deluxetable}

The Magellanic Cloud Emission Line Survey (MCELS) has provided H$\alpha$ $\lambda$6563, 
[\ion{O}{3}] $\lambda$5007, and [\ion{S}{2}] $\lambda\lambda$6716, 6731 line images as 
well as red continuum ($\lambda_c$ = 6850 \AA, $\Delta \lambda$ = 95 \AA) and green continuum 
($\lambda_c$ = 5130 \AA, $\Delta \lambda$ = 155 \AA) images at a $\sim$2$''$
resolution \citep{Smith1999}.  These images have been used to produce continuum-subtracted,
flux-calibrated line images.  For higher-resolution line images, we have also used H$\alpha$,
[\ion{S}{2}], and [\ion{O}{3}] images taken with the MOSAIC camera on the Blanco 4\,m Telescope
at Cerro Tololo Inter-American Observatory (CTIO). The image pixel size is 0.27$''$, and
the seeing measured from the FWHM of stellar image profiles is $\sim$1$''$. % 0.9" - 1.1"
We have carried out a crude flux-calibration for
nebular emission in the 4m MOSAIC H$\alpha$ and [\ion{S}{2}] images by scaling from the 
flux-calibrated MCELS images.  Ten regions of different surface brightnesses without significant
star emission were selected from the H$\alpha$ and [\ion{S}{2}] images, and the scaling is based on
linear regressions between the MCELS fluxes and the MOSAIC counts. These flux-calibrated MOSAIC 
images will be used mainly to search for [\ion{S}{2}]/H$\alpha$ enhancements in 30 Dor B.

In addition to optical images, high-dispersion echelle spectra of
the H$\alpha$ + [\ion{N}{2}] and [\ion{S}{2}] lines reported by
\citet{Chu1992} are re-examined for correlations between high-velocity
shocked gas and diffuse X-ray emission detected in the 2 Ms Chandra  
observation.  

In the Chandra archive, we have found 59 datasets from Advanced CCD Imaging Spectrometer 
\citep[ACIS;][]{Garmire2003} observations of 30 Dor, including the 2 Ms observations from
the T-Rex program and the earlier short exposures.  Only one dataset was obtained with 
the ACIS-S and the rest were obtained with ACIS-I.  
These Chandra datasets, obtained by the Chandra X-ray Observatory, can be
retrieved from \dataset[DOI: 10.25574/cdc.156]{https://doi.org/10.25574/cdc.156}.

The Chandra Interactive Analysis of 
Observations \citep[CIAO;][]{Fruscione2006} version 4.15 and the calibration database (CALDB)
version 4.10.2 were used to process the data, remove the periods of background flares, and 
produce a 2.18 Ms Chandra X-ray image in the $0.5 - 7.0$\,keV band for this paper. 
The image has a pixel size of 0.5$''$ and a resolution of $\sim$1$''$.
The details of reduction and analysis of the Chandra X-ray observations will be reported 
in a separate paper (Ueda, S., et al.\ 2023, in preparation). 

The Australian Square Kilometre Array Pathfinder (ASKAP) has mapped the LMC at 
888 MHz with a beam size of 13.9$'' \times 12.1''$ and a bandwidth of 288 MHz
\citep[Project AS033;][]{Pennock2021}, and 1420 MHz with a beam size of 
9.6$'' \times 7.3''$ and
a bandwidth of 288 MHz (Project AS108).  These high-resolution maps were used 
to derive spectral index and compared with the X-ray image to assess the 
effectiveness of three canonical SNR diagnostics, namely, diffuse X-ray 
emission, nonthermal radio spectral index, and high [\ion{S}{2}]/H$\alpha$ ratios.

For completeness, we have also examined the Spitzer Space Telescope images of 30 Dor B,
particularly the IRAC 8 $\mu$m image that shows polycyclic aromatic hydrocarbon (PAH) 
emission from the partially dissociated regions, and the MIPS 24 $\mu$m image that shows 
thermal dust emission.  These data are from the Spitzer survey of the LMC 
\citep[PI: M.\ Meixner;][]{Meixner2006}.  The resolution of the IRAC 8 $\mu$m image is
$\sim$2$''$ and MIPS 24 $\mu$m $\sim$6$''$.

The above multi-wavelength images are shown in Figure~\ref{fig:multi_band}, which 
includes the 4m MOSAIC images in H$\alpha$, [\ion{S}{2}], and [\ion{O}{3}] lines, 
HST H$\alpha$ image, 4m MOSAIC [\ion{S}{2}]/H$\alpha$ ratio map, Chandra X-ray image
in the 0.5--7.0 keV energy band, ASKAP 888 MHz map, and Spitzer 8 and 24 $\mu$m images.

Finally, for the stellar content of LH99 we use the photometry and spectroscopic 
classifications from the VLT-FLAMES Tarantula Survey \citep{Evans2011,Schneider2018}.
This survey, like the HTTP program, includes both the 30 Dor giant \ion{H}{2}
region and the 30 Dor B region.

%https://archive.stsci.edu/publishing/data-attributions
%We suggest that authors cite the DOI near the end of a section entitled, e.g., "Data" or "Observations."

%https://cxc.cfa.harvard.edu/cda/cdc_doi.html

%\subsection{NANTEN/MAGMA CO Observations}
% NANTEN shows CO, but resolution too low
% MAGMA did not detect CO at 30 Dor

%\subsection{Gamma-Ray Observations}
%Is 30 Dor B a gamma-ray source?

\begin{figure*}
%\epsscale{0.85}
\plotone{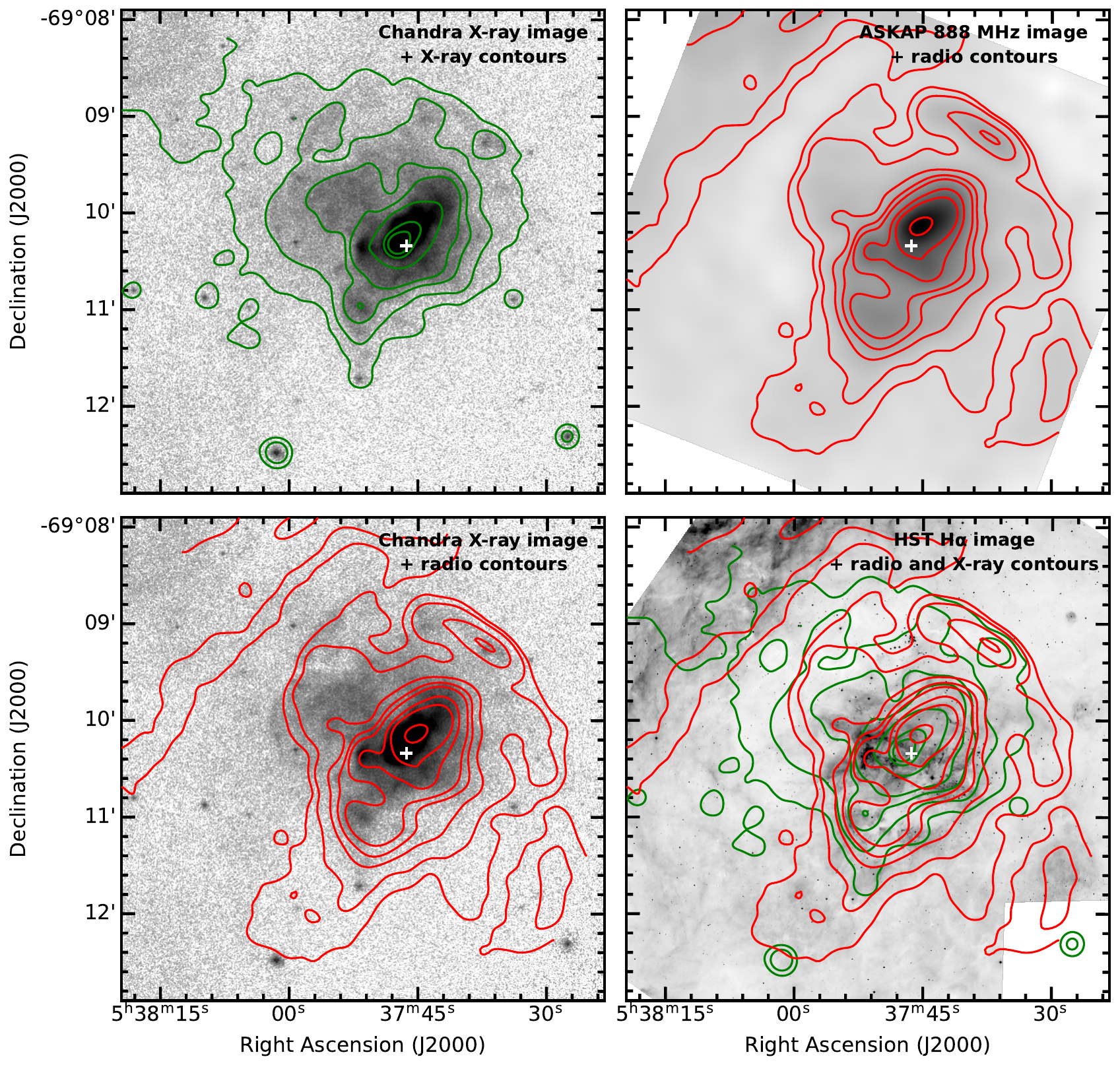}
\caption{Top left: Chandra X-ray image in the 0.5--7.0 keV band shown in logarithm scale. 
The contour levels are $2.3 \times 10^{-9}$, $3.7 \times 10^{-9}$, $8.0 \times 10^{-9}$, 
$1.4 \times 10^{-8}$, $5.0 \times 10^{-8}$, $5.0 \times 10^{-7}$, and $1.0 \times 10^{-6}$ 
counts s$^{-1}$ cm$^{-2}$, extracted from the X-ray image smoothed by a Gaussian filter 
with $\sigma$ = 4$''$. Top right: ASKAP 888 MHz image shown in square root scale.  The contour
levels are 0.007, 0.01, 0.02, 0.03, 0.04, 0.07, 0.1, 0.2 Jy beam$^{-1}$. Bottom left: Chandra X-ray image overplotted with the radio contours from the 
top right panel. Bottom right: HST H$\alpha$ image overplotted with the X-ray contours (green) 
from the top left panel and radio contours (red) from the top right panel. The white ”+” symbol marks
the position of the pulsar PSR J0537$-$6910.}
\label{fig:xray_radio_contour}
\end{figure*}

%======================================================================
\section{Multi-Wavelength View of 30 Dor B} \label{sec:multiwave}  
%======================================================================

Figure~\ref{fig:multi_band} shows multi-wavelength images of 30 Dor B.
While evolved SNRs usually show shell structures in optical emission lines,
30 Dor B does not exhibit any shell structure in the 4m MOSAIC images in 
H$\alpha$, [\ion{S}{2}], and [\ion{O}{3}] lines. The H$\alpha$ and [\ion{S}{2}]
images are qualitatively similar, and the [\ion{O}{3}] image appears more 
amorphous.  The narrow filaments in H$\alpha$ and [\ion{S}{2}] images become
much less conspicuous in the [\ion{O}{3}] line.  

The HST H$\alpha$ image clearly resolves sharp filaments associated with SNR shocks 
that are not discernible in ground-based images. Figure~\ref{fig:30DorB_filaments} 
shows a close-up of the HST H$\alpha$ image of 30 Dor B, and the same image with
visually identified filaments marked.  Long filamentary arcs are present, 
but still no coherent shell structure can be recognized in the HST H$\alpha$ image. 

The Chandra X-ray image in the 0.5--7.0 keV band extracted from 2.18 Ms 
observations shows not only the bright PWN, but also a 
large shell structure about 2.6$'$ ($\sim$40 pc) in diameter.
The PWN is off-center toward the south edge of the shell structure.
The extended diffuse X-ray emission has no obvious optical counterpart, and the X-ray
shell extends well beyond the bright photoionized region of 30 Dor B.  Outside 
the X-ray shell, a faint X-ray halo is detected at least 0.5$'$ further out
from the east shell rim.  The X-ray shell and halo are better seen in 
Figure~\ref{fig:xray_radio_contour}.  In the top left panel of X-ray image
with contours, the X-ray shell is delineated by the lowest contour on the 
west side and the second lowest contour on the east side, while the halo is 
delineated by the lowest contour on the east side.  In the bottom right
panel, the X-ray contours are plotted over the HST H$\alpha$ image, 
clearly showing that the diffuse X-ray emission is much more extended than 
the H$\alpha$ emission region.

The ASKAP 888 MHz map also peaks within the PWN.  In addition, it 
shows an extended component.  As the radio emission from 30 Dor B contains both a 
nonthermal (synchrotron radiation) component from the SNR and a thermal 
(bremsstrahlung radiation) component from the photoionized gas, we 
can compare the radio map to the X-ray and H$\alpha$ images to assess the 
nature of the radio emission.  The top right panel and bottom left panel of 
Figure~\ref{fig:xray_radio_contour} show radio contours over the radio image and
over the X-ray image, respectively.  The main part of the radio emission has many 
correspondences with the X-ray image: (1) both radio and X-ray images exhibit 
prominent emission from the PWN; (2) both exhibit a bright spot 
to the southeast of the pulsar 
wind nebula and appear to be bound by a long H$\alpha$ arc in the HST image; and
(3) the radio image shows a shell structure similar to the X-ray shell with similar
surface brightness variations.  The most obvious differences between the radio and
X-ray images are: (1) radio emission is not detected in the faint X-ray halo region;
(2) the diffuse radio emission patches to the southeast and to the southwest
of the main part have counterparts in the H$\alpha$ but not X-rays.  These southeast and
southwest patches of radio emission and H$\alpha$ emission most likely originate
from diffuse photoionized gas, instead of shocked gas associated with the 30 Dor B
SNR.   

The Spitzer IRAC 8 $\mu$m image shows bright emission from the dark cloud 
in the south part of 30 Dor B, but no morphological feature can be 
unambiguously identified to be associated with SNR shocks.  
The Spitzer MIPS 24 $\mu$m image shows a bright point source in the 
dark cloud and bright diffuse thermal dust emission, but no features are clearly 
associated with SNR shocks.  The bright IR point source, 2MASS J05375027$-$6911071
or MSX LMC888, has been observed with the Spitzer IRS and its spectrum shows nebular
emission lines superposed on dust continuum emission, suggesting the presence of a compact
\ion{H}{2} region ionized by a young massive star \citep{Micelotta2009}, consistent
with the young stellar object classification by \citet{GC2009}. 

%======================================================================
\section{Kinematic Detection of Shocks in 30 Dor B} \label{sec:kinematic}  
%======================================================================

\begin{figure*}
  \centering
  \begin{tabular}{@{}c@{}}
  \hspace{-0.8cm}
    \includegraphics[width=.81\linewidth]{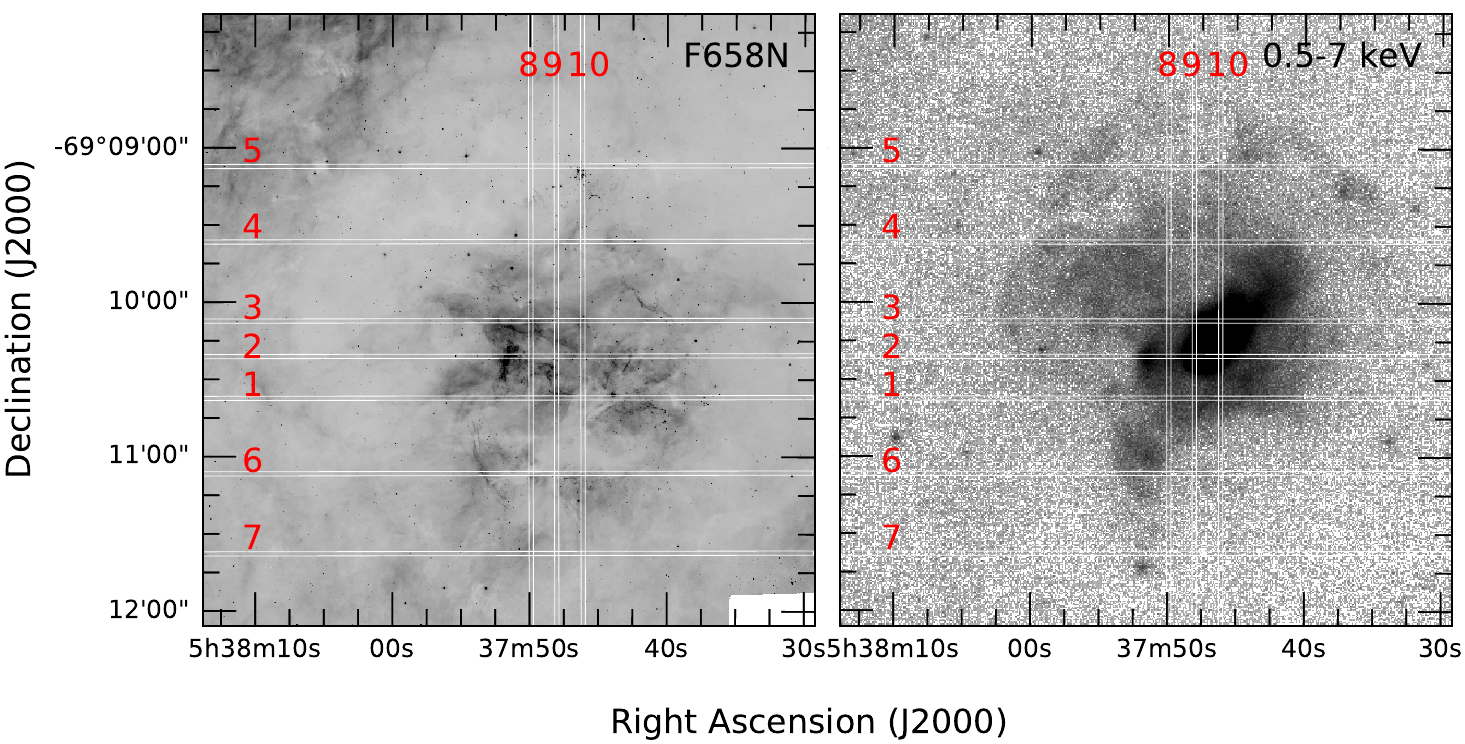} %\\[\abovecaptionskip]
    %\small 
  \end{tabular}
  %\vspace{\floatsep}
  \begin{tabular}{@{}c@{}}
    \includegraphics[width=.8\linewidth]{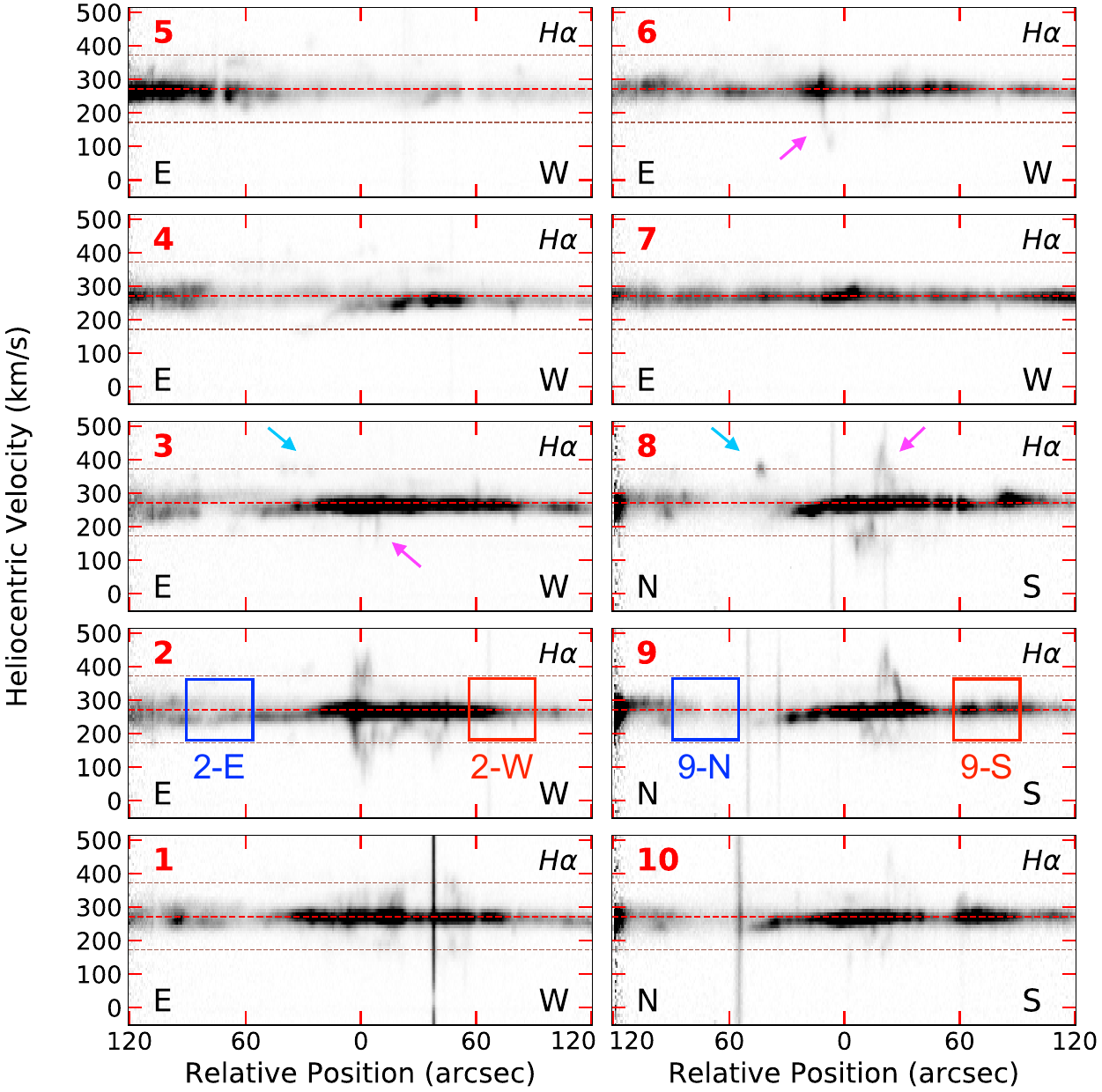} %\\[\abovecaptionskip]
    %\small 
  \end{tabular}
  \caption{Top: CTIO echelle slit positions 1--10 overplotted on the HST F658N (left) and X-ray 
  Chandra 0.5-7.0 keV (right) images. The X-ray image is displayed in logarithmic scale. Bottom: 
  CTIO 4m H$\alpha$ echellograms displayed with {\it spatial scale and extent matching the slit positions
  plotted in the top panels}. The slit numbers are marked in the upper-left corners in red.  A thick dashed
  line is plotted at the bulk ISM velocity of 272 km s$^{-1}$, and two auxiliary thin dashed lines at 
  $\pm$100 km s$^{-1}$ offsets, respectively. Magenta arrows point at examples of high-velocity features 
  emanating from the quiescent ISM velocity, and cyan arrows point at examples of discrete 
  high-velocity cloudlets.  Pixels affected by cosmic ray hits have been fixed by interpolation between 
  adjacent unaffected pixels. The H$\alpha$ line profiles in boxes 2--E (blue), 2--W (red), 9--N (blue), and 9--S (red) are presented in Figure \ref{fig:30DorB_echelle_spectra_FWHM}.  }
\label{fig:30DorB_echelle_1}
\end{figure*}

%\begin{figure*}
%  \centering
%  \begin{tabular}{@{}c@{}}
%  \hspace{-0.8cm}
%    \includegraphics[width=.81\linewidth]{30DorB_echelle_positions} %\\[\abovecaptionskip]
%    %\small 
%  \end{tabular}
%  %\vspace{\floatsep}
%  \begin{tabular}{@{}c@{}}
%    \includegraphics[width=.8\linewidth]{30DorB_echelle_spectra_low_vel} %\\[\abovecaptionskip]
%    %\small 
%  \end{tabular}
%  \caption{Top: CTIO echelle slit positions overplotted on the optical HST F658N (left) and X-ray %Chandra 0.5-7 kev (right) images. The X-ray image is displayed in the log scale. Bottom: CTIO echelle %H$\alpha$-line spectra showing the structure of the bulk ionized gas. The slit numbers (red) are marked %in the upper-left corners and correspond to those in the top panel. }
%\label{fig:30DorB_echelle_1}
%\end{figure*}

\begin{figure*}
  \centering
  \begin{tabular}{@{}c@{}}
  \hspace{-0.8cm}
    \includegraphics[width=.81\linewidth]{30DorB_echelle_positions} %\\[\abovecaptionskip]
    %\small (a)
  \end{tabular}
  \vspace{\floatsep}
  \begin{tabular}{@{}c@{}}
    \includegraphics[width=.8\linewidth]{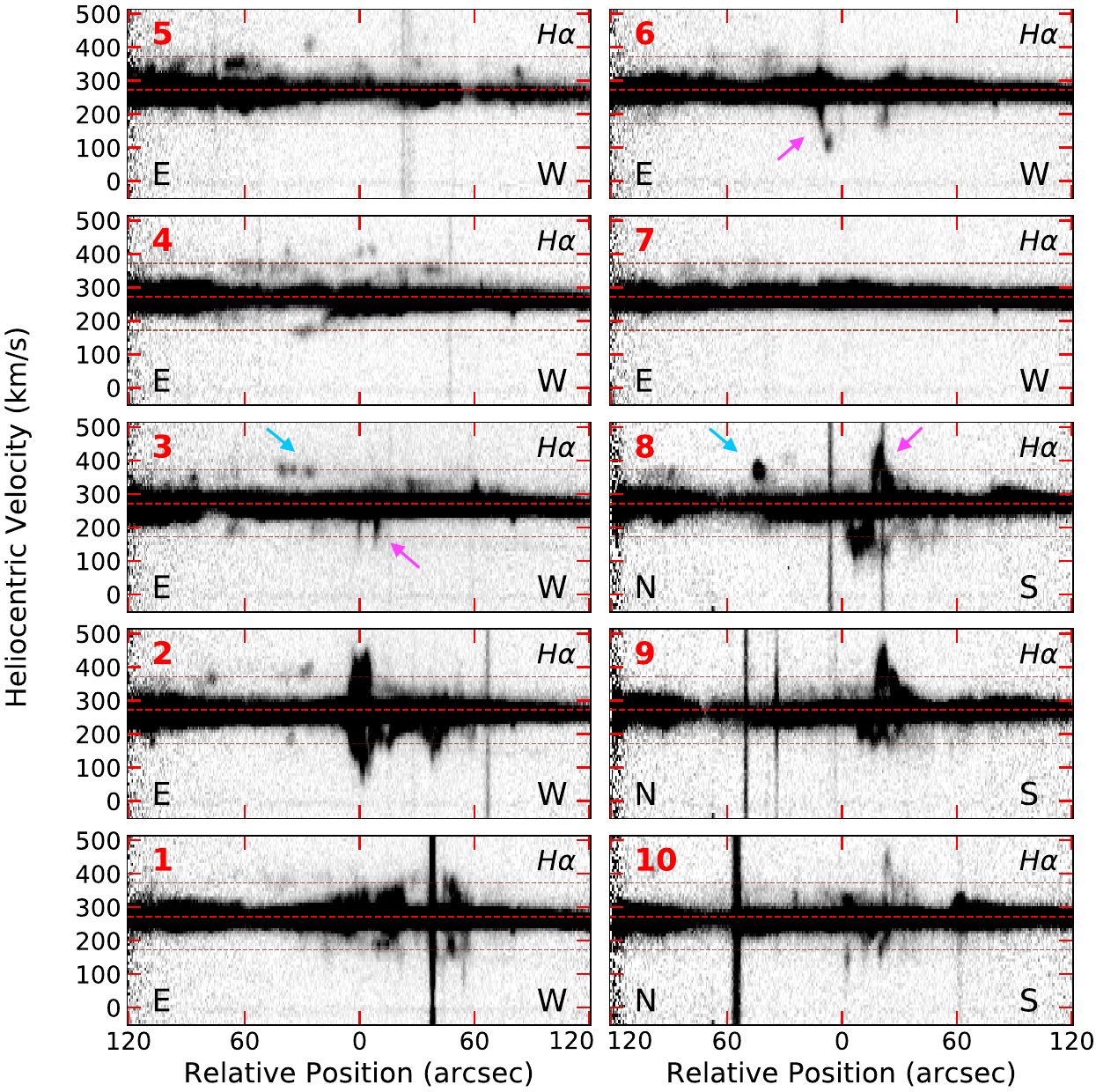} %\\[\abovecaptionskip]
    %\small (b)
  \end{tabular}
  \caption{Same as Figure \ref{fig:30DorB_echelle_1}, but the echellograms in the bottom panels
  are displayed with a different intensity scale to accentuate the faint high-velocity features.}
\label{fig:30DorB_echelle_2}
\end{figure*}

\begin{figure}  
% \hspace{-0.8cm}
\epsscale{1.}
\plotone{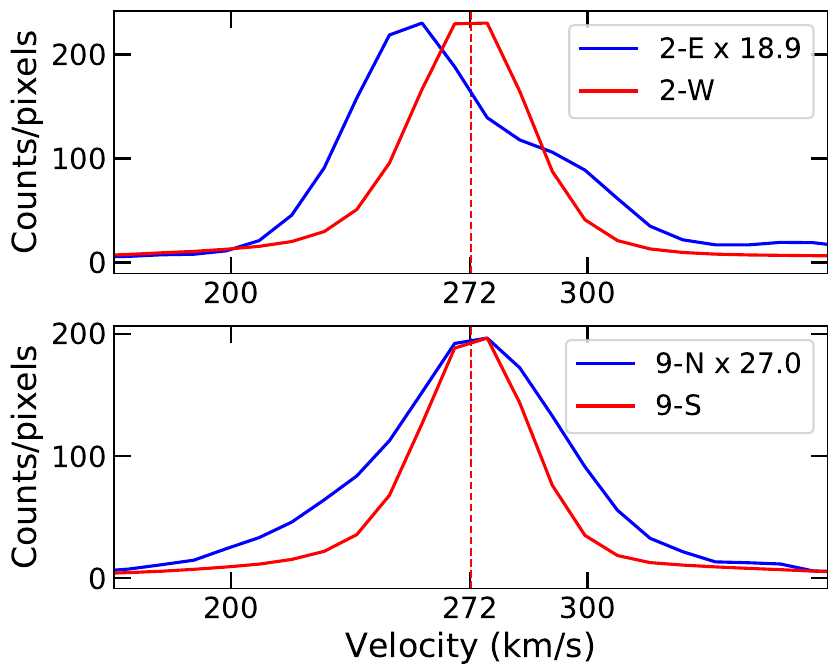} 
\caption{Top panel: H$\alpha$ velocity profiles extracted from boxes
2--E (blue) and 2--W (red) along slit position 2. Bottom panel:
H$\alpha$ velocity profiles extracted from boxes 9--N (blue)and 9--S (red)
along slit position 9.  The box positions are marked in 
Figure \ref{fig:30DorB_echelle_1}.  The fainter profile has been multiplied by
a factor given in the upper right corner of the panel to match the peak of the 
brighter profile. The instrumental FWHM measured from the telluric lines is 
$\sim$18 km\,s$^{-1}$; thus even the narrow profiles are resolved. }
\label{fig:30DorB_echelle_spectra_FWHM}
\end{figure}

%\begin{figure*}
%  \centering
%  \begin{tabular}{@{}c@{}}
%  \hspace{-0.8cm}
%    \includegraphics[width=.81\linewidth]{30DorB_echelle_positions} %\\[\abovecaptionskip]
    %\small (a)
%  \end{tabular}
%  \vspace{\floatsep}
%  \begin{tabular}{@{}c@{}}
%    \includegraphics[width=.8\linewidth]{30DorB_echelle_spectra_high_vel} %\\[\abovecaptionskip]
    %\small (b)
%  \end{tabular}
%  \caption{Same as Figure \ref{fig:30DorB_echelle_1}, but the bottom panel is for showing the faint high-velocity features. }
%\label{fig:30DorB_echelle_2}
%\end{figure*}

The kinematic properties of the 30 Dor B region have been studied with
long-slit high-dispersion echelle spectroscopic observations \citep{Chu1992}.
The spectra were compared with those of \ion{H}{2} regions, SNRs, superbubbles,
and the 30 Dor giant \ion{H}{2} region. They concluded that the 30 Dor B SNR 
was $1.5'\times1.2'$ in size and that high-velocity interstellar features 
were present to the north and to the east of the SNR.  Given the extended
diffuse X-ray emission revealed by the deep Chandra observation, we ought
to re-examine the velocity structures, assess their relationship 
with the diffuse X-rays, and investigate whether these features are associated 
with a single SNR or multiple SNRs in 30 Dor B.

For easier visualization of correlations between the kinematic features and
the morphological features, we mark the slit positions on the HST H$\alpha$
image as well as the Chandra X-ray image, and re-present the long-slit 
echellograms from \citet{Chu1992} with spatial extent and scale matching 
the slit positions marked in the images.  The echellograms are displayed 
in two intensity scales, accentuating the bright low-velocity ISM in 
Figure~\ref{fig:30DorB_echelle_1}
and the faint high-velocity material in 
Figure~\ref{fig:30DorB_echelle_2}.
The radial velocities are expressed in heliocentric velocities 
throughout this paper.

The bulk ionized gas in 30 Dor B is best seen in slit positions 1--3.  
The dense unperturbed photoionized gas in 30 Dor B has a constant radial 
velocity at 272$\pm$5 km\,s$^{-1}$.  To the south, slit positions 6 and 7 
show that the bulk gas has a similar constant radial velocity.  This constant
velocity is most likely the systemic velocity of the general ISM in 30 Dor B.
In slit positions 1--5, to the east of 30 Dor B the velocity profiles 
become broader and the central velocity shows significant variations, 
indicating a more perturbed velocity structure.  
In slit positions 8--10, the bright ionized gas in 30 Dor B shows a 
consistent constant radial velocity of $\sim$272 km\,s$^{-1}$, but the 
faint exterior to the north appears much more perturbed. 
Examples of narrow-velocity profiles from quiescent regions 
(positions 2-W and 9-S marked in Fig.~\ref{fig:30DorB_echelle_1}) and 
broad-velocity profiles from perturbed regions (positions 2-E and 9-N
marked in Fig.~\ref{fig:30DorB_echelle_1}) are shown in
Figure~\ref{fig:30DorB_echelle_spectra_FWHM}.
%The more perturbed velocity structure exterior to the northeast quadrant 
%of 30 Dor B is understandable, as this is the ``strait'' between 30 Dor
%and 30 Dor B where the abundant high-velocity gas in 30 Dor would
%inevitably spill over. 

It has been observed in the LMC that confirmed SNRs show shell expansion 
velocities over 100 km\,$^{-1}$ or shocked material at $>$100 km\,s$^{-1}$ 
velocity offsets, while photoionized \ion{H}{2} regions are quiescent without 
any material accelerated to $>$100 km\,s$^{-1}$ \citep{CKSNR1988,CKHII1994}.
Thus, in Figures~\ref{fig:30DorB_echelle_1} and \ref{fig:30DorB_echelle_2}, we 
plot a thick dashed line at the bulk ISM velocity of 272 km\,$^{-1}$, and 
two auxiliary thin dashed lines at $\pm$100 km\,$^{-1}$ offsets, respectively,
to visualize interstellar features with velocity offsets greater than 
100 km\,s$^{-1}$ that are most certainly associated with SNRs.

Two types of high-velocity features are seen in the long-slit echellograms.
The first type appears emanating from the quiescent interstellar velocity
and extending to velocity offsets of 100--200 km\,s$^{-1}$; some of them
curve back to the interstellar velocity, forming loop-like structures in the 
echellograms, while others end at highest velocity offsets, forming spikes or
arc-like structures in the echellograms.  Examples of these features are
marked by magenta arrows in Figures \ref{fig:30DorB_echelle_1} and 
\ref{fig:30DorB_echelle_2}.  The second type appears as discrete 
high-velocity cloudlets without any connection to the quiescent
interstellar velocity. Examples of these features are marked by cyan arrows
in Figures \ref{fig:30DorB_echelle_1} and \ref{fig:30DorB_echelle_2}.  
Based on these different types of features, 
\citet{Chu1992} argued that the first type of features was frequently seen 
in SNRs and thus they must be associated with the 30 Dor B SNR, but was 
uncertain about the second type of features and suggested that they were 
high-velocity interstellar clouds.

It can be seen in Figures~\ref{fig:30DorB_echelle_1} and \ref{fig:30DorB_echelle_2} 
that these two types of high-velocity features are apparently associated with 
different interstellar environments. The first type is superposed on the 
brightest parts of the photoionized ISM in 30 Dor B, while the second type 
is observed outside the brightest parts of 30 Dor B.
The surface brightness of the photoionized gas is proportional to the electron
density squared; thus, a high surface brightness means a high gas density.
Therefore, the first type of high-velocity features is associated with shocks
in an extended dense medium, while the second type of high-velocity feature is
associated with discrete cloudlets in a tenuous medium.

\begin{figure}
\epsscale{1.15}
\hspace{-0.6cm}
\plotone{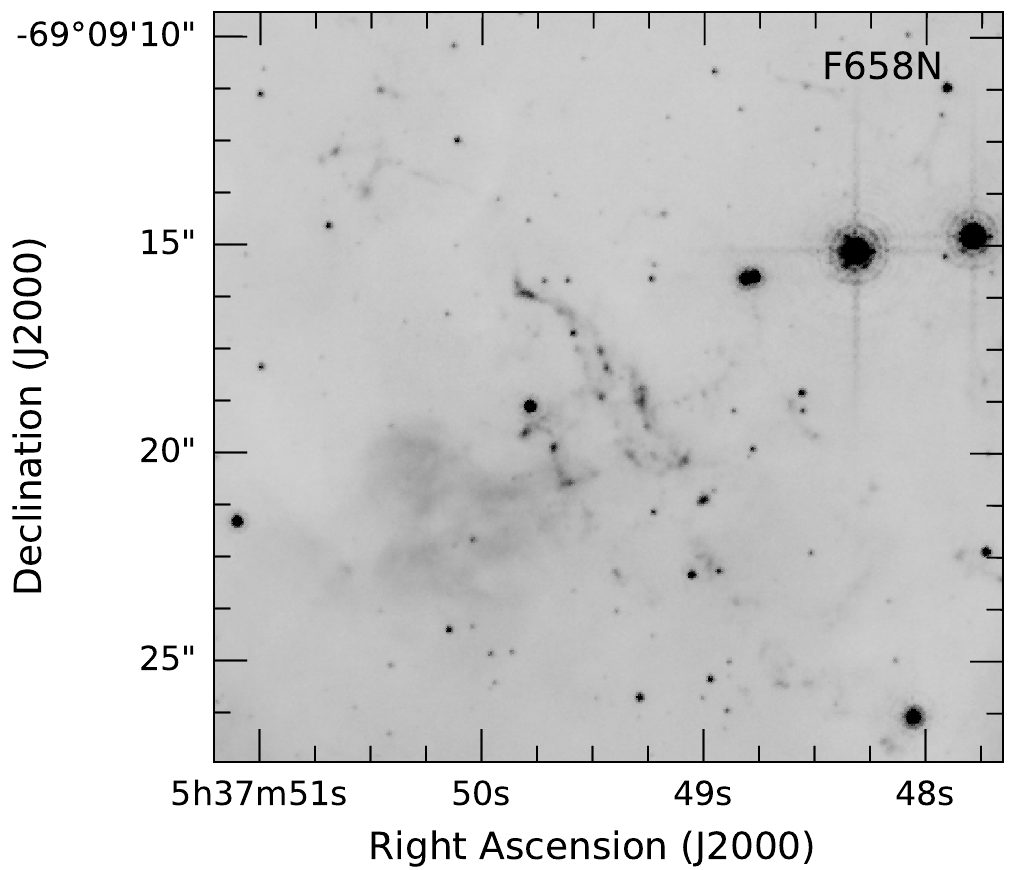}
\caption{Close-up of shocked cloudlets.}
\label{fig:cloudlets}
\end{figure}

Examining the locations of the high-velocity cloudlets and the
distribution of diffuse X-ray emission in Figure~\ref{fig:30DorB_echelle_2},
we can see that most of the high-velocity cloudlets are projected within 
the 2.6$'$-diameter X-ray shell.  There are still some faint high-velocity cloudlets
projected within the faint X-ray halo outside the X-ray shell, but these
cloudlets have lower velocity offsets, generally $<$100 km\,s$^{-1}$ from
the quiescent interstellar velocity.  Among these high-velocity cloudlets,
only the brightest one in slit position 8 has an obvious counterpart in the HST 
H$\alpha$ image, as shown in Figure~\ref{fig:cloudlets}.  This 6$''$ 
high-velocity feature is resolved into knots strung along filaments.
This morphology is seen in shocked interstellar cloudlets in core-collapse
SNRs as well as circumstellar medium in Type Ia SNRs 
\citep[see examples given in Figure 8 of][]{Li2021}.  Thus, these 
high-velocity features are most likely shocked and shredded cloudlets.

\begin{figure*}
\epsscale{1.15}
\hspace{-0.5cm}
\plotone{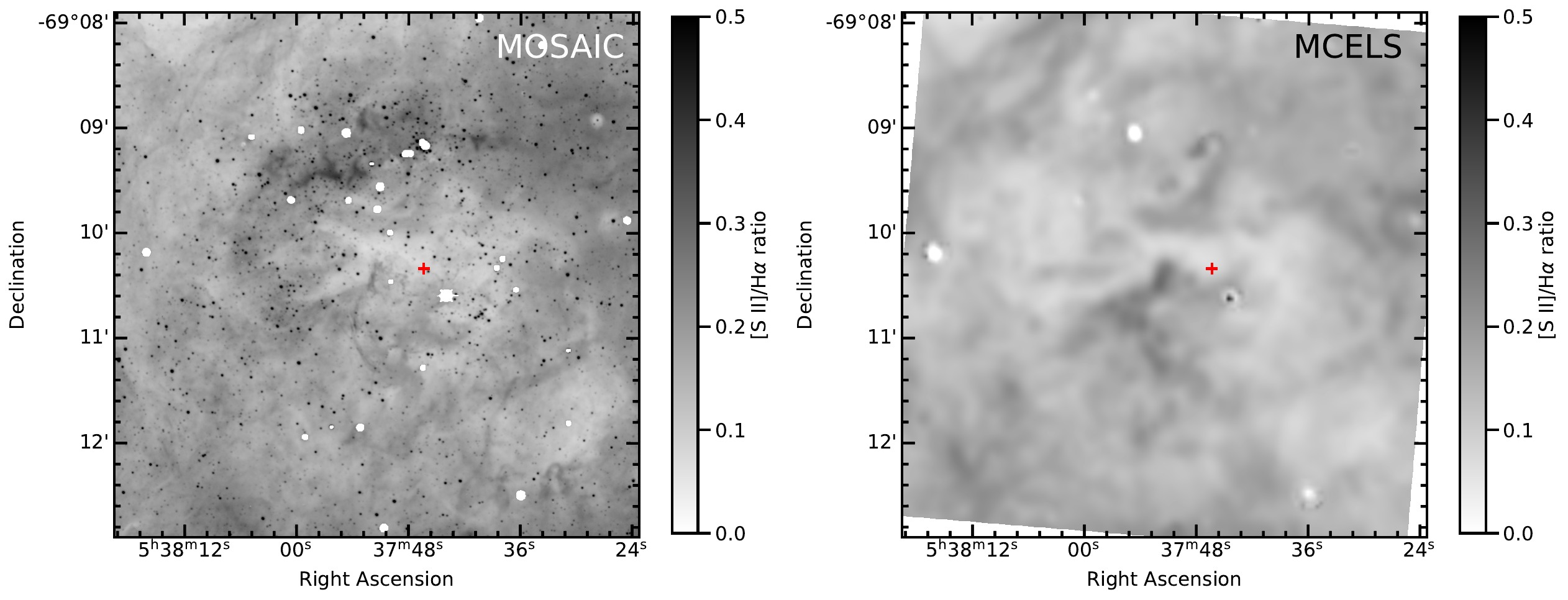}
\caption{[\ion{S}{2}]/H$\alpha$ ratio maps derived from the 4m MOSAIC images (left) and MCELS images
(right). The white spots in the MOSAIC map are caused by saturated star images. The red ”+” symbol marks the position of the pulsar.}
\label{fig:sii_ha_raio}
\end{figure*}

\section{Discussion}

The complex X-ray morphology and kinematic properties of 30 Dor B
raise several questions: 
How effective are the SNR diagnostics at different wavelengths? 
Has 30 Dor B resulted from more than one SN explosion?  
Can the mass of the SN progenitor be assessed?
The first question is addressed in Section~\ref{sec:SNR_diag}
and the latter two in Section~\ref{sec:progenitors}.

\subsection{Effectiveness of SNR Diagnostics} \label{sec:SNR_diag}

Conventionally, SNRs are identified and confirmed by three 
signatures: diffuse X-ray emission, nonthermal radio emission,
and high [\ion{S}{2}]/H$\alpha$ ratios.  The multi-wavelength
views of 30 Dor B reveal that some diagnostics appear more effective 
than the others in a complex environment.  The X-ray image is 
most effective in revealing shock-heated hot gas. It has been 
shown that XMM-Newton observations with $\sim$20 ks exposure 
can detect diffuse X-ray emission from an SNR in an ISM with 
densities as low as 0.01 H-atom cm$^{-3}$ \citep{Ou2018}.

Optical line emission from SNRs are more dependent on the 
physical conditions of the ambient medium.  For a young Type Ia 
SNR in a partially neutral medium, its collisionless shocks 
produce hydrogen recombination lines, and its optical spectrum 
is dominated by Balmer lines \citep{Chevalier1980, Heng2010}.
For a young core-collapse SNR, if the SN progenitor was a massive 
O star and its strong stellar wind has already cleared out the
ambient ISM, the SNR shock would advance in a low-density ionized medium
and the post-shock material would be too hot and tenuous to emit
detectable hydrogen Balmer lines; however, if remnant dense cloudlets 
remain in the surroundings of a SN progenitor, the SNR shocks into
the dense cloudlets will become radiative early and emit optical lines
and show high [\ion{S}{2}]/H$\alpha$ ratios, for example, the SNR N63A
\citep{Levenson1995}.
If a SN progenitor was an early B star without strong stellar wind,
its dense ambient medium will cause the SNR shock to become radiative 
soon and produce bright line emission with high [\ion{S}{2}]/H$\alpha$ 
ratios, for example, the SNR N49 \citep{Bilikova2007}.

In the case of 30 Dor B, the SNR shocks advance into a complex ISM with
both ionized gas and dusty neutral material.  The co-location of the SNR
and the ISM, as well as the similarity in their systemic velocities, 
suggest that the SNR is indeed interacting with the \ion{H}{2} region in 
30 Dor B.  Because of the complex and nonuniform surrounding medium, no
coherent SNR shell can form within the \ion{H}{2} region, although 
fragmented partial shell structures are seen, as marked in 
Figure \ref{fig:30DorB_filaments}.
On a larger scale, we see the X-ray shell that has no optical counterpart.
This X-ray shell must be caused by the SNR shock advancing into an ionized 
low density medium, and the shocks are not radiative yet.  Thus, optical
line images are not as effective as X-ray images in revealing SNR shocks.

Spectroscopic detection of shocked dense gas provides another useful
diagnostic of SNR shocks.  While the extended diffuse X-ray emission
region does not show any optical shell counterpart, high-velocity 
cloudlets are detected in the H$\alpha$ line.  Similar high-velocity 
cloudlet features are commonly seen in 30 Dor; however,
the photoionized \ion{H}{2} region in 30 Dor is so bright that the 
faint high-velocity cloudlets cannot be identified morphologically 
in the H$\alpha$ images \citep{Chu1994}.  In 30 Dor B, for the first 
time, a high-velocity cloudlet finds a morphological counterpart 
in an H$\alpha$ image!  As shown in Figure~\ref{fig:cloudlets}, the bright 
high-velocity cloudlet along slit position 8 is resolved in the HST H$\alpha$
image into shreds of filaments and knots. The presence of high-velocity
cloudlets in the diffuse X-ray emission region indicates that small
dense cloudlets are embedded in a low-density ISM.

The [\ion{S}{2}]/H$\alpha$ ratio has been commonly used as a diagnostic
in optical SNR surveys.  
SNRs in an isolated environment can have [\ion{S}{2}]/H$\alpha$ 
ratios approaching or greater than 1; however, SNRs superposed on \ion{H}{2} 
regions will have this diagnostic diluted by the emission from the 
photoionized gas.  Thus, a threshold of [\ion{S}{2}]/H$\alpha$ $\ge$ 0.45 
has been used to identify extragalactic SNRs \citep[e.g.,][]{MF1997}.  
The [\ion{S}{2}]/H$\alpha$ ratio map of 30 Dor B (Fig.~\ref{fig:multi_band})
shows low values in the bright photoionized \ion{H}{2} region, and high
values to the north of 30 Dor B.  While there is some correspondence between
the extended diffuse X-ray emission and the enhanced [\ion{S}{2}]/H$\alpha$
ratio, the details are confused by photoionization structures on the surface of
dark clouds, large density variations, and complex stellar and interstellar 
environments, rendering the [\ion{S}{2}]/H$\alpha$ an ineffective SNR diagnostic
in 30 Dor B.

[\ion{S}{2}]/H$\alpha$ ratio maps derived from the 4m MOSAIC and the MCELS
images are shown in Figure \ref{fig:sii_ha_raio}. The different appearances of
these two maps amply illustrate the dilution of SNR emission by the background
\ion{H}{2} region emission. The MCELS images have a 2$''$ pixel size that is almost
ten times the width of the filaments resolved in the HST H$\alpha$ image.
The MOSAIC images have a higher resolution and a smaller pixel size, 0.27$''$, thus
the MOSAIC [\ion{S}{2}]/H$\alpha$ ratio map shows higher values and more fine 
structures.  The [\ion{S}{2}]/H$\alpha$ ratio appears enhanced along 
an arc from the north to the east roughly along the rim of the X-ray shell and halo.
As this arc is also just outside the photoionized gas, it is not clear whether 
the enhanced [\ion{S}{2}]/H$\alpha$ is caused by shocks or ionization fronts. 

\begin{figure*}
\epsscale{1.15}
\plotone{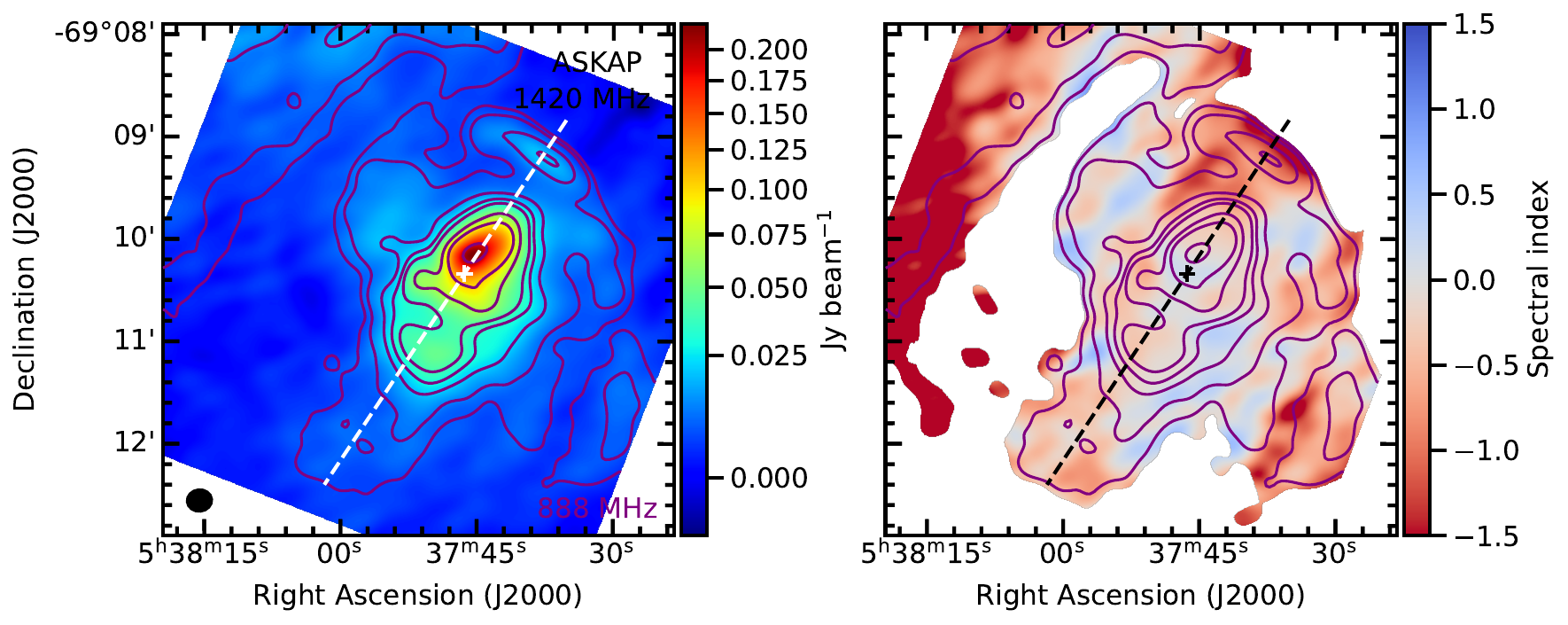}
\caption{Left panel: ASKAP 1420MHz map smoothed to match the angular resolution 
of the 888 MHz map. Right panel: spectral index map.
Both panels are overplotted with contours from 888 MHz map shown in Figure \ref{fig:xray_radio_contour}. In the spectral index map, only the regions with 
888 MHz flux density $> 15 \sigma$ are shown, where 
$\sigma \sim 4\times 10^{-4}$ Jy beam$^{-1}$. 
The ”+” symbol marks the position 
of the pulsar. The dashed lines connect the radio/X-ray arc, PWN, 
pulsar, and the bright patches of emission inside and outside
the X-ray-emitting regions. The very steep negative spectral 
index near the east edge of the map is part of 30 Dor giant \ion{H}{2} region.}
\label{fig:N157B_spectral_index}
\end{figure*}

The comparisons among H$\alpha$, radio 888 MHz, and X-ray images described in
Section~\ref{sec:multiwave} indicate that 888 MHz radio maps can be as 
effective as X-ray images in detecting SNRs in an ionized low density medium, 
where optical H$\alpha$ emission cannot be detected.
When multi-frequency radio observations are available, the spectral index $\alpha$
in $S_\nu \propto \nu^{\alpha}$, where $S_\nu$ is the flux density and $\nu$ the 
frequency, can be derived to assess whether the radio emission is thermal or nonthermal.
Normally, thermal emission has $\alpha$ $\sim -0.1$ for an optically thin case and 
approaches 2 for an optically thick case, while $\alpha$ has a steep negative value
for nonthermal emission.  However, 30 Dor B is complicated by the superposing thermal
emission from the photoionized {\ion{H}{2}} region and the flat spectral index of 
the synchrotron radiation from its PWN.
Figure~\ref{fig:N157B_spectral_index} shows the 1420 MHz map convolved to 
match the resolution of the 888 MHz map, and a spectral index map derived from 
these two maps of 30 Dor B.
As the two frequencies are very close and the spectral index from the faint regions can be
quite uncertain, we choose to present spectral index only within regions that are brighter
than the lowest contour that delineate the SNR.  This lowest contour corresponds to
$\sim 18\sigma$ detection at 888 MHz.
In the spectral index map, the radio-brightest region is associated with the pulsar wind
nebula and shows a rather flat spectral index,
fluctuating between $\sim$0.2 and $\sim -0.2$.  This is consistent with the previous 
measurement of $-0.19$ determined from 3.5 and 6 cm observations \citep{Lazendic2000}.
This radio-bright region is superposed on a bright \ion{H}{2} region with very nonuniform
surface brightness, which complicates the interpretation of the spectral index variations
in the radio-brightest central region of 30 Dor B.  It is impossible to disentangle the
synchrotron radiation from the thermal emission, unless the radio maps are available 
over a larger range of frequencies with $\sim$1$''$ resolution that matches the 
ground-based H$\alpha$ image resolution.

Steep negative spectral indexes, $<$$-1$, are observed in the X-ray shell, particularly 
along the northwest rim where a bright radio arc is present and coincides with an X-ray arc.
It is interesting to note that the elongated radio emission in 30 Dor B and its vicinity seems
to align with the pulsar and the elongated outflow of the pulsar wind. To the northwest end
is the center of the radio/X-ray arc, while to the southeast exist a bright 
radio/X-ray spot bordered by a sharp H$\alpha$ arc in the HST image and a purely radio 
extension outside the SNR with spectral indexes 
$\sim -0.5$. The alignment of these features is illustrated by the dashed line drawn 
in each panel of Figure~\ref{fig:N157B_spectral_index}.   It is tempting to suggest 
these features may be associated with an outflow or a jet from the pulsar region.  
Nevertheless, it should be pointed out that outside the X-ray region there are two 
patches of radio emission, the aforementioned southeast extension 
and a detached southwest patch, and both have amorphous diffuse H$\alpha$ emission counterparts.
The nature of these two patches of radio/H$\alpha$ emission is uncertain.
Future radio observations with wider frequency coverage and higher angular resolution
may help determine the nature of these features more accurately.

\begin{figure*}
\epsscale{1.}
\plotone{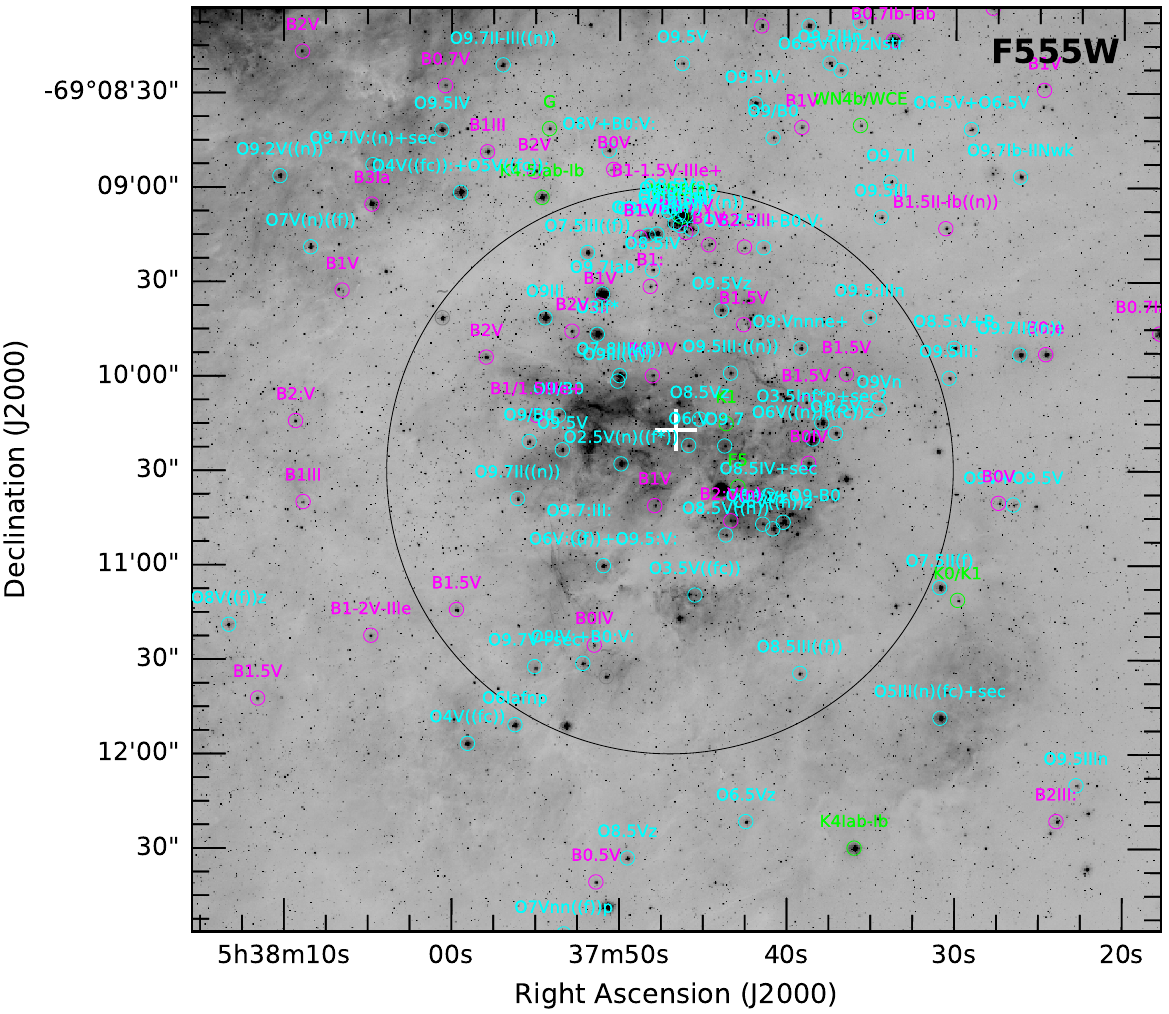}
\caption{The HST F555W image of the SNR 30 Dor B and its environment.  The spectral
types of massive stars from the VLT-FLAMES Tarantula Survey \citep{Evans2011} are
marked, with O and B stars in cyan and magenta, respectively.
Late-type supergiants and Wolf-Rayet stars are marked in green. 
The 16 ms pulsar in the SNR is marked with a white cross. The 3$'$-diameter
circle marks a rough boundary of the OB association LH99 \citep{Lucke1970}.}
\label{fig:N157B_spec_types}
\end{figure*}

\begin{figure*}
\epsscale{1.1}
\plottwo{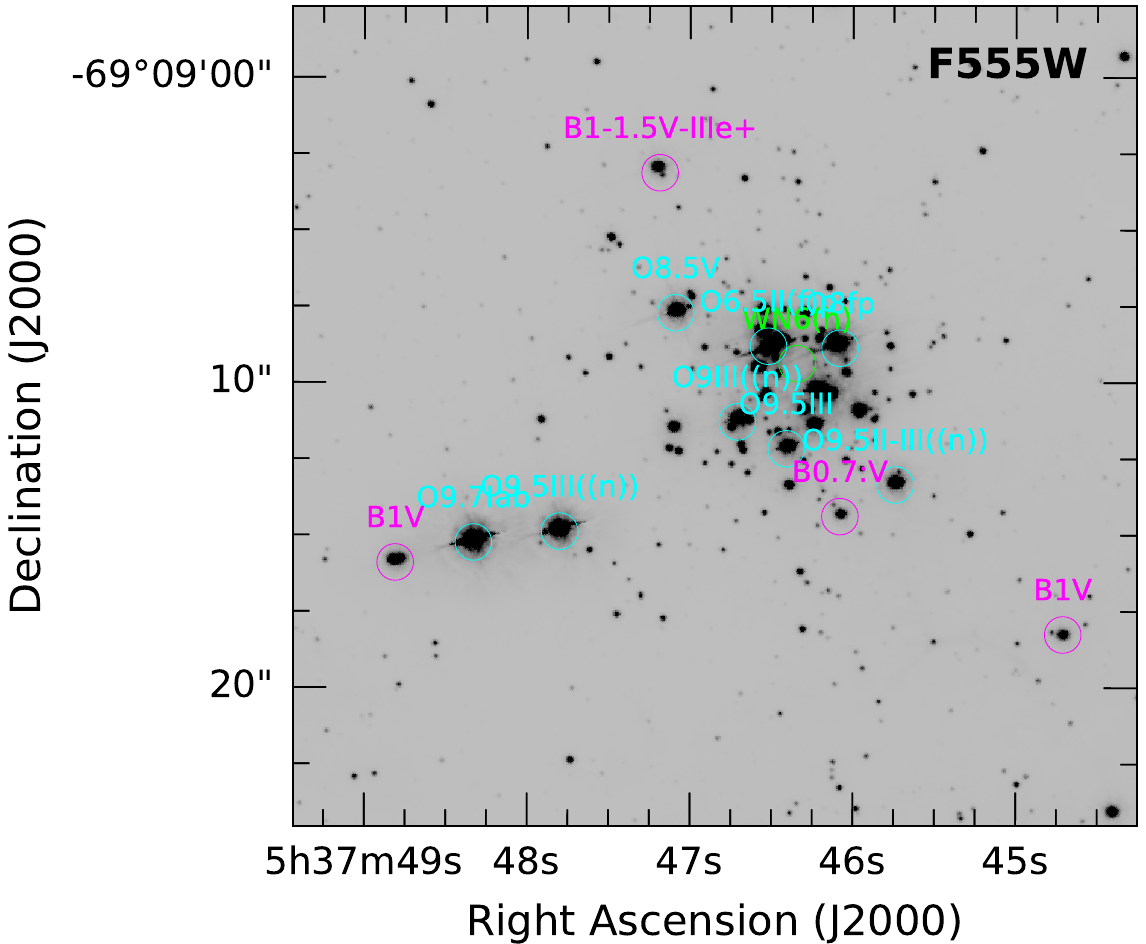}{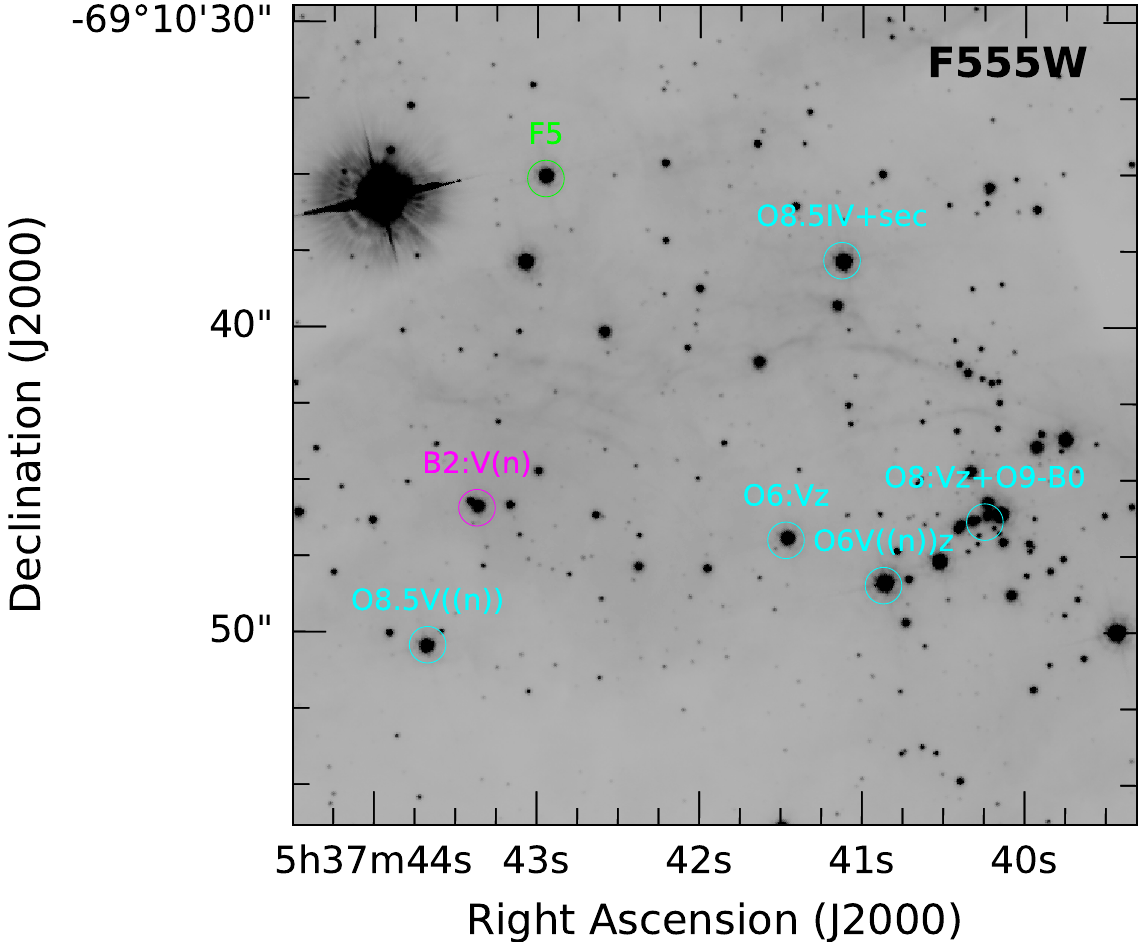}
\caption{ Close-up images for two crowded regions in Figure \ref{fig:N157B_spec_types}. The
small cluster in the left panel is TLD 1.}
\label{fig:N157B_spec_types_crowded}
\end{figure*}

\subsection{SN progenitors of 30 Dor B} \label{sec:progenitors}

30 Dor B can be divided into three zones according to the distribution 
of X-ray emission and kinematic properties of the dense ionized gas.
The innermost zone, $\sim$20 pc (1.3$'$) in size, shows SNR shocks impacting on the 
dense ionized gas in the bright \ion{H}{2} region, in addition to the pulsar 
wind nebula.  Incidentally, the region corresponds to the 1.5$'$ $\times$ 1.2$'$
SNR boundary suggested by \citet{Chu1992}. 
The second zone is the X-ray shell that is about 40 pc (2.6$'$) across and extends
much beyond the bright \ion{H}{2} region, where high-velocity cloudlets
with velocity offsets greater than 100 km\,s$^{-1}$ are observed.
The third zone is the faint X-ray halo exterior to the X-ray shell, where 
high-velocity cloudlets are observed but the velocity offsets are generally
much less than 100 km\,s$^{-1}$. The distinct properties of these three
zones are usually associated with SNRs at different evolutionary stages 
and/or interstellar environments. It is unlikely that one single SN is
responsible for all these observed properties.

To examine the underlying stellar population in and around 30 Dor B, we  
have used the Magellanic Cloud Photometric Survey 
\citep[MCPS,][]{Zaritsky2004} to produce color--magnitude diagrams (CMDs)
overplotted with stellar evolutionary tracks or isochrones, similar to the 
analysis of SNR B0532$-$67.5 \citep{Li2022}.  The distributions of stars 
in the CMDs do not indicate a simple single burst of star formation; furthermore,
there exist very massive stars whose masses cannot be determined reliably from 
photometry at optical wavelengths.  Thus, we resort to the spectroscopic 
VLT-FLAMES Tarantula Survey \citep[VFTS,][]{Evans2011}. 

The VFTS has been used to make spectroscopic classifications of massive stars, 
assess their masses and ages, and determine the star formation history in
NGC\,2070 (aka 30 Dor) and NGC\,2060 (aka 30 Dor B or LH99) by \citet{Schneider2018}.
They report that star formation in the past 8--10 Myr proceeded
from north to northwest of 30 Dor toward southeast, along the
molecular ridge between 30 Dor and 30 Dor B, and that the age distribution
of the stars in 30 Dor B peaks at 4.4 Myr with a median of 5.7 Myr.  
They find that to the north of LH99, at 5h37m46s, $-$69$^\circ$09$'$10$''$ (J2000) 
is the small association TLD 1, which is possibly 3.3--3.5 Myr old.  
Therefore, the star formation in 30 Dor B is by no means single-burst, and it 
is thus impossible to associate the SN progenitor with a particular stellar age group.

To illustrate the complexity of stellar population in 30 Dor B, we have
marked the spectral types of massive stars in 
Figure~\ref{fig:N157B_spec_types}.  Close-ups of TLD 1 and another crowded
region are presented in Figure~\ref{fig:N157B_spec_types_crowded}.
It is evident that LH99 in 30 Dor B contains O2.5 to O3 main sequence stars
whose initial masses are greater than 100 $M_\odot$.  The upper mass limit
for massive stars to end in neutron stars is 25--30 $M_\odot$ \citep{Heger2003}.  
Thus, the progenitor of the pulsar in 30 Dor B could not have formed at the same 
time as these very massive stars, and no useful constraint on the SN progenitor's
mass can be derived.  

While we cannot assess the SN progenitor mass, we may assess whether the
three zones with distinct X-ray and kinematic properties resulted from 
three different events.  The central region of 30 Dor B associated with 
bright X-ray emission (including the PWN) and shocked ISM
at velocity offsets greater than 200 km\,s$^{-1}$ is likely the most
recent SN event from a progenitor that had been a late-O main sequence star.
The age of this SNR can be approximated by the spin down time of its pulsar,
5000 yr \citep{Marshall1998}. If this SN event has also produced the X-ray 
shell, the expected expansion velocity of the X-ray shell, approximated
by 0.5(radius/age), would be 4000 km\,s$^{-1}$.  Such a high velocity
would produce post-shock plasma at keV temperatures,
which is not supported by the electron temperature of $kT \sim 0.3$ keV 
derived from modeling of the X-ray spectra (Ueda et al.\ 2023, in preparation).
Therefore, this X-ray shell cannot be formed by the same SN event as the pulsar,
and it must have resulted from an earlier SN explosion.
The faint X-ray halo is much larger and the shocked cloudlets have even smaller
velocity offsets from the ISM, these properties are similar to those seen in 
large-scale diffuse X-ray emission whose SNRs can no longer be unambiguously 
identified or traced. Considering that star formation has been on-going for the 
past several Myr, it is not surprising that multiple SNe have occurred within 
the vicinity of 30 Dor B.
We conclude that the observed physical properties of 30 Dor B need at least 
two SN events.  The faint X-ray halo may belong to a large-scale background
produced by previous SN events.

%The age of the X-ray shell can be estimated from a simple approximation of 
%0.5(radius/expansion velocity). Assuming that the expansion velocity is 
%100 km\,s$^{-1}$, the radius of 20 pc requires a shell age of $\sim10^5$ yr.
%This age is much greater than the 30 Dor B pulsar's spin-down time, 5000 yr
%\citep{Marshall1998}. 
%Even if the expansion velocity is 500 km\,s$^{-1}$, the age of the X-ray
%shell is still much larger than the pulsar's spin-down time.

%======================================================================
\section{Summary} \label{sec:Summary}  
%======================================================================

30 Dor B, also cataloged as NGC\,2060, is located at the southwest of
the giant \ion{H}{2} region 30 Dor.  30 Dor B contains the OB association 
LH\,99, the \ion{H}{2} region photoionized by LH\,99, and 
MCSNR J0537$-$6910.  The \ion{H}{2} region is superposed by dark clouds,
making it very difficult to study the boundary and structure of the
SNR.  Recent HST H$\alpha$ images from the HTTP program (PI: E.\ Sabbi) 
have resolved the narrow filaments associated with SNR shocks, but no 
shell structure is observed.  The deep 2 Ms Chandra X-ray image from the 
T-Rex program (PI: L.\ Townsley) has detected not only the pulsar wind
nebula, but also an extended X-ray shell $\sim$40 pc in size and an additional
faint halo.  The X-ray shell and halo are much more extended than the 
\ion{H}{2} region of 30 Dor B. The ASKAP 888 MHz map of 30 Dor B shows 
counterparts to all X-ray emission features except the faint X-ray halo.

We have re-analyzed the high-dispersion long-slit echelle observations of
\citet{Chu1992}.  In addition to the quiescent component of unperturbed 
ISM at 272$\pm$5 km\,s$^{-1}$, we find two types of high-velocity features
offset by more than 100 km\,s$^{-1}$.  The first type appears to emanate
from the quiescent interstellar velocity and extend to velocity offsets of 
100--200 km\,s$^{-1}$, and these features are associated with dense ISM.
In contrast, the second type appears as discrete high-velocity cloudlets without
any connection to the quiescent interstellar velocity, and these features
are associated with less dense ISM.  We further find that high-velocity 
features of the first type are associated with the central X-ray-bright 
region, while those of the second type are within the extended X-ray shell.
There are high-velocity cloudlets within the faint X-ray halo, but their
velocity offsets from the quiescent interstellar velocity are usually 
$<$100 km\,s$^{-1}$.

The multi-wavelength observations of 30 Dor B are used to evaluate the 
effectiveness of the three common SNR diagnostics: diffuse X-ray emission, 
[\ion{S}{2}]/H$\alpha$ ratio, and nonthermal radio spectral index. 
In 30 Dor B's complex environment where a SNR is superposed on a photoionized 
\ion{H}{2} region dissected by dark clouds, diffuse X-ray emission offers 
the most effective diagnostic of SNRs, while the [\ion{S}{2}]/H$\alpha$ 
ratio is the least effective diagnostic of SNR shocks.  Radio emission is 
more effective than optical emission in detecting SNRs, but the confusion caused
by the superposed thermal emission renders the spectral index less effective in
diagnosing SNR accurately. 

The spectroscopic VFTS survey of massive stars in 30 Dor B is
used to deduce that star formation has been going on for the past
8--10 Myr and that the most massive stars in 30 Dor B have initial masses
greater than 100 $M_\odot$ \citep{Schneider2018}.  Based on this information,
we conclude that the SN progenitor mass cannot be assessed from the
complex star formation history and stellar population.  Adopting the pulsar's
spin-down time of 5000 yr as the SNR age, the X-ray shell would have to be
expanding at 4000 km\,s$^{-1}$ to be produced by the same SN explosion.  
Such a fast expansion would produce post-shock plasma temperature incompatible
with that indicated by X-ray spectra.  We conclude that at least two SN events 
had occurred in 30 Dor B and the faint X-ray halo may require additional 
previous SN events.

\acknowledgments
We thank the anonymous referee for prompt reviews and constructive suggestions to 
improve this paper. We also thank Dr.\ John Raymond for useful discussions on shock.  
YHC and CJL acknowledge the support of the NSTC 
grants 110-2112-M-001-020 and 111-2112-M-001-063 
and SU acknowledges the NSTC grant NSTC 111-2112-M-001-026-MY3 from the National Science and 
Technology Council of Taiwan. 
\\

%https://ui.adsabs.harvard.edu/help/faq/
This research has made use of NASA’s Astrophysics Data System Bibliographic Services,
%http://simbad.u-strasbg.fr/simbad/sim-basicIdent=m33&submit=SIMBAD+search
and the SIMBAD database,
operated at CDS, Strasbourg, France \citep{Wenger2000}.
% https://cxc.harvard.edu/cda/acknowledgment.html
This paper has used data obtained from the Chandra Data Archive and the Chandra Source Catalog, and software provided by the Chandra X-ray Center (CXC) in the application packages CIAO and Sherpa.
%https://research.csiro.au/casda/about/
%https://www.atnf.csiro.au/research/publications/Acknowledgements.html
%https://noirlab.edu/science/about/scientific-acknowledgments#ctio
This paper has also used observations at Cerro Tololo Inter-American Observatory at NSF’s NOIRLab, which is managed by the Association of Universities for Research in Astronomy (AURA) under a cooperative agreement with the National Science Foundation.
% Publications that use data from ASKAP or the Murchison Radio-astronomy Observatory should include a statement as follows: 
Also used in this paper are ASKAP 888 MHz and 1420 MHz maps.
The Australian SKA Pathfinder (ASKAP) is part of the Australia Telescope National Facility which is managed by CSIRO. Operation of ASKAP is funded by the Australian Government with support from the National Collaborative Research Infrastructure Strategy. ASKAP uses the resources of the Pawsey Supercomputing Centre. Establishment of ASKAP, the Murchison Radio-astronomy Observatory and the Pawsey Supercomputing Centre are initiatives of the Australian Government, with support from the Government of Western Australia and the Science and Industry Endowment Fund. We acknowledge the Wajarri Yamatji people as the traditional owners of the Observatory site.
%https://irsa.ipac.caltech.edu/data/SPITZER/docs/spitzermission/publications/ackn/
This work is based in part on observations made with the Spitzer Space Telescope, which was operated by the Jet Propulsion Laboratory, California Institute of Technology under a contract with NASA.\\

%https://journals.aas.org/facility-keywords/
\emph{Facilities:} CXO (ACIS), HST (ACS, WFC3), CTIO Blanco (MOSAIC), CTIO Curtis Schmidt, Spitzer (IRAC, MIPS), ASKAP\\

\emph{Software:} SAOImage DS9 \citep{Joye2003}, Astropy \citep{Astropy2013, Astropy2018, Astropy2022}, Matplotlib \citep{Hunter2007}, NumPy \citep{vanderWalt2011, Harris2020}, SciPy \citep{Virtanen2020}, CIAO \citep{Fruscione2006},
XSPEC \citep{Arnaud1996}

\end{document}